\documentclass{article}

\usepackage[figuresright]{rotating}
\usepackage{graphics}
\usepackage{amsmath}
\usepackage{amsfonts}
\usepackage{natbib}

\title{Modelling overdispersion heterogeneity \\ in differential expression analysis using mixtures}

\author{Elisabetta Bonafede$^{1}$,Franck Picard$^{2}$,St\'{e}phane Robin$^{3,4}$, and Cinzia Viroli$^{1}$ \\
$^{1}$ Department of Statistical Sciences, University of
  Bologna, Italy\\
  $^{2}$ Laboratoire de Biom\'etrie et Biologie
  \'Evolutive, \\ UMR CNRS 5558 Univ. Lyon 1, F-69622 Villeurbanne, France, \\
   $^{3}$  AgroParisTech, UMR 518 MIA, Paris, France,\\
   $^{4}$  INRA, UMR 518 MIA, Paris, France}

\begin{document}
\maketitle

\begin{abstract}
Next-generation sequencing technologies now constitute a method of choice
to measure gene expression. Data to analyze are read counts,
commonly modeled using Negative Binomial distributions. A relevant
issue associated with this probabilistic framework is the reliable
estimation of the overdispersion parameter, reinforced by the
limited number of replicates generally observable for each gene. Many
strategies have been proposed to estimate this parameter, but when
differential analysis is the purpose, they often result in procedures
based on plug-in estimates, and we show here that this discrepancy
between the estimation framework and the testing framework can lead to
uncontrolled type-I errors. Instead we propose a mixture
model that allows each gene to share information with other genes that exhibit similar variability. Three consistent statistical tests are
developed for differential expression analysis. We show that the proposed method improves the
sensitivity of detecting differentially expressed genes with respect to the common procedures, since  it is the best one in reaching the nominal value for the
first-type error, while keeping elevate power. The method is finally illustrated on prostate cancer RNA-seq data.
\end{abstract}

\maketitle

\section{Introduction}\label{sec:intro}
Massive parallel sequencing has deeply changed our understanding of
gene expression thanks to a higher resolution \citep{soon13,
wang2009rna}.  From the analysis point of view, NGS experiments
provide discrete \emph{read counts} assigned to target genome regions
measuring the expression level or the abundance of the target
transcript. When the purpose of the assay is to perform
\emph{differential analysis}, that is comparing the counts of a given
regions between conditions, the statistical task is then to provide an
appropriate model to account for biological and technical variations,
as well as a testing framework to test the hypothesis of no
difference. Here we deal with the case where regions of interest are given {\sl a priori}, contrarily to analysis where the regions themselves have to be discovered \citep{FSH14}.
Generalized linear models based on count distributions now constitute
a consensus framework for the analysis, with the original Poisson distribution \citep{marioni08, wang10} being replaced by the Negative Binomial model \citep{robinson2008, anders2010differential,robinson2010edger}. Indeed, the simplest choice of the Poisson distribution was rapidly identified as the cause of uncontrolled first-type errors, due to a poor adjustment to the larger observed variability compared with the equal mean-variance specification of the Poisson model (see, for a discussion, \cite{anders2010differential}). Since then, the correct modeling and estimation of this observed overdispersion has been a key issue in differential analysis.

Taking perspective from our past experience in micro-array analysis, the proper modeling of the dispersion parameter has long been a subject of debate in differential analysis, with a difficult trade-off between a common variance for every genes and gene-specific variances. Given the limited number of replicates, the first strategy provides robust estimates, but the testing procedure lacks of power and the model is not realistic, whereas the second is more sensitive at the price of increased first-type errors. Actually, the debate is still ongoing with the Negative Binomial framework, but the problem is much more difficult to solve due to this complex (and unknown) mean-variance relationship.

Several contributions have been proposed to find a trade-off between the common overdispersion and the gene-specific overdispersion frameworks. \cite{robinson2008} addressed the problem in a multi-step procedure called \emph{edgeR}. They first proposed to estimate a common dispersion parameter for all genes expressed as a quadratic combination of the mean, and then, by making use of a weighted likelihood procedure, they provide an estimation of each dispersion parameter as a weighted combination of the common and of the individual ones, assuming empirical weights. Then an approximation is introduced in order to develop an exact test.
\cite{anders2010differential} proposed to use a mean-dependent local regression to smooth the gene-specific dispersion estimates, related to the idea that genes that share a similar mean expression level have also a similar variance, and therefore they can contribute to the estimation of the respective parameters. The method is implemented in the \emph{DESeq} R package. \cite{WWW13} introduced an empirical Bayes shrinkage approach choosing a log-normal prior distribution on the dispersion parameters and therefore imposing a negative binomial likelihood. Then the estimations are plugged-in the Wald statistic to perform the statistical test. The method is implemented in the \emph{DSS} R library.  In the \cite{hk10} proposal, the dispersion is iteratively estimated using the quasi-likelihood approach. A comparison of all these existing approach has been illustrated in \cite{YHV13}, where a new strategy based on the method of the moments is employed in order to get reliable dispersion parameters estimation.

Despite the rapidly increasing diffusion of these statistical procedures, also thanks to the availability of the well documented Bioconductor packages such as \emph{edgeR}, \emph{DESeq} and \emph{DSS}, the estimation of the dispersion in NGS data remains a crucial and tricky issue because of the limited number of available observations for each gene. Moreover, less attention has been focused on the consistency between the estimation and the testing frameworks. Indeed, an important drawback of using plug-in estimators is that the expected variations of the test statistics are no longer controlled under the null hypothesis, which may result in an un-controlled level of the test. We will illustrate this point in a simulation study, by showing that most proposed methods do not reach the nominal level of the test, whereas it is precisely what is expected to be controlled when performing standard hypothesis testing.

Our contribution is to explore and discuss a mixture model approach \citep{mclachlan2000finite, Fraley2002} based on the idea of sharing information among genes that exhibit similar dispersion. More specifically, mixtures of negative binomial distributions are investigated as a way to get more accurate estimation for the dispersion parameter of each gene, exploiting also the information provided by the others. Such an approach has already be considered in the same context for the differential analysis of microarray data \citep{DRD05}. A consistent statistical testing procedure is then developed within the unified model based clustering framework. The proposed method improves the sensitivity of detecting differentially expressed genes with small replicates, and we show that our method controls the nominal level of the test by simulation.

The novel method will be described in Section
\ref{sec:meth}. We will derive three statistical tests and describe
the procedure for performing the differential analysis in Section
\ref{sec:statistics}. We will show through a large simulation study in
Section \ref{sec:sim} that the proposed statistical test procedure
outperforms the mostly used strategies in the literature, because it
is the best one in reaching the nominal value for the first-type
error, while keeping elevate power, thus indicating its inferential
reliability. The method will be applied on prostate cancer data in Section
\ref{sec:appl}. A final discussion is presented in Section
\ref{sec:dis}.

\section{The proposed method}\label{sec:meth}

\subsection{The data problem}
Suppose we observe the read counts of $p$ genes in $d$ biological conditions. For each condition, assume that $n_j$ replicates are available, with $j=1,\ldots,d$.
Without loss of generality, here we consider $d=2$. We denote $Y_{ijr}$ the random variable that expresses the read counts, say $y_{ijr}$, mapped to
gene $i$ ($i$=$1,...,p$), in condition $j$, in sample $r$ ($r$= $1,...,n_j$). Let $\textbf{Y}_i$ be the random vector of length $n=\sum_{j=1}^dn_j$ denoting the
expression profile of a gene. Let $\textbf{y}_i$ be the observed value. Analogously, $\textbf{y}_{ij}$, is the vector of length $n_j$ containing the observed gene
profile under condition $j$.
Let us consider the Negative Binomial distribution (NB), $NegBin(\lambda,\alpha)$ with parameters $\lambda$, $\alpha > 0$, expectation $\lambda$ and variance
$\lambda \left(1+\frac{1}{\alpha} \lambda \right)$  (see, for instance, \cite{hilbe2011}), so that $\lambda$ is the location parameter and $\alpha$ is the
dispersion parameter. For a random variable $Y \sim NegBin(\lambda,\alpha)$ we have
\begin{equation}\label{eq:densNB}
f(y | \lambda, \alpha)=  \binom{y+\alpha-1}{\alpha-1}  \left( \frac{\lambda}{\lambda + \alpha} \right)^y \left( \frac{\alpha}{\lambda + \alpha} \right )^{\alpha}
\mbox{.}
\end{equation}
By assuming that the read counts can be described by the distribution in (\ref{eq:densNB}), ideally we would fit the model
$$Y_{ijr} \sim NegBin(\lambda_{ij},\alpha_i)$$ for each gene and then
we would test the null hypothesis $H_0: \lambda_{i1}=\lambda_{i2}$. In
most experiments, replicates are too few to accurately estimate both
parameters for each gene (usually less than 20).  To overcome the
limitedness of the available observations, we propose to aggregate
information of genes sharing similar variability thought a mixture
model that jointly consider all the genes. The proposal constitute a
trade-off between the gene specific variance model and the one with
common dispersion parameter.

\subsection{The model}
An interesting characteristic of the NB parametrization in
(\ref{eq:densNB}) is that it can be derived from a Poisson-Gamma mixed
distribution. More precisely, if the random variable $U_{ijr}$ follows
the Gamma distribution with unit mean $$U_{ijr} \sim
Gamma(\alpha_i,\alpha_i),$$ and the random variable $Y_{ijr}$
conditional on $U_{ijr}=u_{ijr}$ follows the Poisson
distribution $$Y_{ijr}|U_{ijr}=u_{ijr} \sim Poisson(\lambda_{ij}
u_{ijr}),$$ then unconditionally $Y_{ijr}$ is distributed according to
the NB, $Y_{ijr} \sim NegBin(\lambda_{ij},\alpha_i)$. This means that
the parameter $\lambda_{ij}$ of the Poisson component rules the
expectation of the negative binomial distribution, and the parameter
$\alpha_i$ of the gamma component controls the heterogeneity, thus
allowing overdispersion.

Instead of fitting $p$ different NB models, we consider a unified
mixture model approach for all the genes. In order to develop it,
notice that the data can be viewed in a multilevel structure, where
the first-level units are the replicates ($r=1, ..., n_{j}$), at the
second level we have the conditions ($j=1, ..., d$) and at the third
level there are the genes ($i=1, ..., p$). By considering the
replicates independent draws within and between conditions we have
\begin{eqnarray}
f(\textbf{y}_i)=\prod_{j=1}^d f(\textbf{y}_{ij})=\prod_{j=1}^d
\prod_{r=1}^{n_j}f({y}_{ijr})\mbox{.}
\end{eqnarray}

We consider a mixture model for the units at the third level of this
hierarchical structure. More precisely, we assume that the
heterogeneity component $\textbf{u}_i$ contains $n$ independent
replicates with distribution given by a mixture of $K$ gamma
components:
\begin{eqnarray}
f(\textbf{u}_i)=\sum_{k=1}^K w_k f_k(\textbf{u}_i)= \sum_{k=1}^K w_k
\prod_{j=1}^d
\prod_{r=1}^{n_j}Gamma({u}_{ijr};\alpha_k,\alpha_k)\mbox{,}
\end{eqnarray}
It is possible to show that if $Y_{ijr}|U_{ijr}=u_{ijr} \sim
Poisson(\lambda_{ij}u_{ijr})$ then $\textbf{Y}_{i}$ follows a mixture of NB
distributions:
\begin{eqnarray}\label{eq:mod}
f(\textbf{y}_i)=\sum_{k=1}^K w_k f_k(\textbf{y}_i)= \sum_{k=1}^K w_k
\prod_{j=1}^d
\prod_{r=1}^{n_j}NegBin({y}_{ijr};\lambda_{ij},\alpha_k)\mbox{.}
\end{eqnarray}
Proof is given in the Appendix.

In the mixture model (\ref{eq:mod}) the $p$ genes are clustered
according to their variability only. Conditionally on each group $k$,
the density of the generic read count $y_{ijr} $ is a negative
binomial with gene-wise and condition-wise mean equal to
$\lambda_{ij}$ and variance equal to
$\lambda_{ij}\left(1+\frac{1}{\alpha_k} \lambda_{ij} \right)$.

\subsection{Model estimation}
Let $\boldsymbol\Theta= \{\lambda_{ij},w_k,\alpha_k\}_{i=1,...p;
  j=1,...,d; k=1,...,K}$ be the whole set of model parameters. The
log-likelihood of the model is given by
\begin{eqnarray}\label{Ltheta}
\ln L(\theta)= \ln \prod_{i=1}^p \sum_{k=1}^K w_k \prod_{j=1}^d
\prod_{r=1}^{n_j} NegBin({y}_{ijr};\lambda_{ij},\alpha_k)\mbox{.}
\end{eqnarray}
A direct maximization of (\ref{Ltheta}) is not analytically possible,
but the maximum likelihood estimates can be derived by the EM
algorithm \citep{dempster1977}. The algorithm alternates between the
expectation and the maximization steps until convergence, with the aim
of maximizing the conditional expectation of the so-called complete
log-likelihood given the observable data. The complete log-likelihood
is the joint log-density of the observable data and of the missing
data of the model.  In the proposed mixture model, the hidden data
consists of two sets of latent variables, which are (1) the latent
variable $u$ deriving by considering the NB distribution as a
gamma-poisson mixed process and (2) the latent allocation $K$-vector,
$\textbf{z}_i$, that contains the 1 if the gene belongs to the group
$k$ and zero otherwise. The complete log-likelihood can be defined as

\begin{eqnarray}\label{Lcomplete}
\ln L_c(\theta) &=& \ln f(\textbf{y},\textbf{u},\textbf{z})= \ln
\prod_{i=1}^p f(\textbf{y}_i|\textbf{u}_i)
f(\textbf{u}_i|\textbf{z}_i)f(\textbf{z}_i)\nonumber \\ & =& \ln
\prod_i \prod_j \prod_r f(y_{ijr}|u_{ijr}) f(y_{ijr} {u_{ijr}}
|\textbf{z}_i)f(\textbf{z}_i),
\end{eqnarray}
where $f(y_{ijr}|u_{ijr})=Poisson(y_{ijr};
\lambda_{jr}u_{ijr})$, $f({u_{ijr}}|{z}_{ik}=1)=Gamma({u_{ijr};
}\alpha_k,\alpha_k)$ and $f(\textbf{z}_i)$ is the multinomial
distribution
$$
{f(\textbf{z}_i)} = \prod_{k=1}^K w_k^{z_{ik}}.
$$

 In the EM algorithm we maximize the conditional expectation of the
 complete density given the observable data, using a fixed set of
 parameters $\boldsymbol\Theta'$:
\begin{eqnarray}\label{EM}
&& \arg \max_{\boldsymbol\Theta}E_{\textbf{z},\textbf{u}|\textbf{y}; \boldsymbol\Theta'}\left[
 \log f(\textbf{y},\textbf{u},\textbf{z}|\boldsymbol\Theta)\right],
\end{eqnarray}
which leads to iterating the E and M steps until convergence.
The details of the algorithm are described in the Appendix. The EM algorithm has been implemented in R code and a CRAN package
will be available soon.

\section{Differential analysis}\label{sec:statistics}

In this section, the procedure to identify the genes ($i= 1, ..., p $)
that differentially express under two ($j=1, 2$) different biological
conditions is described.  This aim can be accomplished in different
ways: one could be interested in evaluating the equality between the
two population means, or in checking wether their ratio is equal to 1,
or wether the log-ratio is zero. The three scenarios can be
represented by the following null hypotheses:
\begin{enumerate}
\item ``Difference'': $ H_0: \lambda_{i1}- \lambda_{i2} = 0 $
\item ``Ratio'': $ H_0: \frac{\lambda_{i1}}{\lambda_{i2}} = 1 $
\item ``Log Ratio'': $ H_0: \ln\frac {\lambda_{i1}}{\lambda_{i2}} =
  \ln (\lambda_{i1})- \ln (\lambda_{i2}) = 0 $
\end{enumerate}

For each case, we can evaluate a test-statistic based on the EM
estimates. Since the EM estimators are maximum likelihood estimators,
the test-statistics are asymptotically distributed according to the
standard Gaussian under the null hypothesis:
\begin{enumerate}
\item For the ``Difference'' test statistic: $
  \frac{\widehat{\lambda}_{i1}-
    \widehat{\lambda}_{i2}}{\sqrt{Var(\widehat{\lambda}_{i1}-
      \widehat{\lambda}_{i2})}} | H_0 \rightsquigarrow N(0,1) $,
\item for the ``Ratio'' test statistic: $
  \frac{\frac{\widehat{\lambda}_{i1}}{\widehat{\lambda}_{i2}}-1}{\sqrt{Var
      \left(\frac{\widehat{\lambda}_{i1}}{\widehat{\lambda}_{i2}}\right)}}
  | H_0 \rightsquigarrow N(0,1) $,
\item for the ``Log Ratio'' test statistic: $ \frac{\ln
  \widehat{\lambda}_{i1}- \ln \widehat{\lambda}_{i2}}{\sqrt{{Var}(\ln
    \widehat{\lambda}_{i1}- \ln \widehat{\lambda}_{i2})}} | H_0
  \rightsquigarrow N(0,1) $,
\end{enumerate}
where $\widehat{\lambda}_{i1}$ and $\widehat{\lambda}_{i2}$ are the EM-estimators.

\subsubsection*{The ``Difference'' statistical test}
\ \ \ For computing the test statistic we need to estimate the variance at
the denominator of the test statistics.  First of all we notice that
${Var}(\widehat{\lambda}_{i1}- \widehat{\lambda}_{i2}) =
{Var}(\widehat{\lambda}_{i1}) +{Var}(\widehat{ \lambda}_{i2}) -2 Cov
(\widehat{\lambda}_{i1},
\widehat{\lambda}_{i2})={Var}(\widehat{\lambda}_{i1}) +{Var}(\widehat{
  \lambda}_{i2})$ since the covariance is zero by the model
assumptions. The specific variances $Var(\widehat{\lambda}_{ij})$ are
a function of $y_{ijr}$. In particular, $\widehat{\lambda}_{ij} =
\frac{\sum_{r=1}^{n_j} y_{ijr}}{n_j}$ and denoting $\sum_r y_{ijr} $
as $ y_{ij+}$
\begin{eqnarray}\label{var.y}
Var(\widehat{\lambda}_{ij}) = \frac{1}{n_j^2} Var(y_{ij+}). \end{eqnarray}
The variance $Var(y_{ij+})$ can be computed by observing that the
replicates $y_{ijr}$, with $r=1,\ldots,n_j$ are independently
distributed according to a mixture of NB, so that $Var(y_{ij+})= n_j
Var(y_{ijr})$ and for mixture models the general formula for the
variance holds
\begin{eqnarray}
Var(y_{ijr})&=& E[Var(y_{ijr}|z_{ik})]+Var[E(y_{ijr}|z_{ik})] \nonumber \mbox{,}
\end{eqnarray}
where $Var[E(y_{ijr}|z_{ik})] = 0$ because the expectation is not
component varying; as regards $E[Var(y_{ijr}|z_{ik}))]$ we
considered the conditional expectation given the observed data because
of the multilevel structure of the data, and therefore
\begin{eqnarray}\label{var.y2}
Var(y_{ijr})&=&E_{\textbf{z}_i |\textbf{y}_i}[Var(y_{ijr}|z_{ik}=1)]=
\widehat{\lambda}_{ij}\left(1+ \widehat{\lambda}_{ij} \sum_k
 \frac{f(z_{ik}|\textbf{y}_i)}{\alpha_k} \right).
\end{eqnarray}
This formula enlightens the effect of the mixture model we propose:
the over-dispersion term is a weighted average of the (estimated)
over-dispersion terms $\lambda_{ij}/\alpha_k$ one would get in each
component of the mixture. These terms are weighted according to the
posterior probability for observation $i$ to belong to each component
$k$: $f(z_{ik}|\textbf{y}_i)$. \\

\subsubsection*{The ``Ratio'' statistical test}
\ \ \ As for the computation of $
Var\left(\frac{\widehat{\lambda}_{i1}}{\widehat{\lambda}_{i2}} \right)
$ we can use Delta method (\cite{van2000,cox1990}), and by simple
computations we get:
$$ Var \left(
\frac{\widehat{\lambda}_{i1}}{\widehat{\lambda}_{i2}}\right) \approx
\frac{Var(\widehat{\lambda}_{i1})}{E(\widehat{\lambda}_{i2})^2} +
\frac{E(\widehat{\lambda}_{i1})^2}{E(\widehat{\lambda}_{i2})^4}
Var(\widehat{\lambda}_{i2}) $$ All the needed quantities can be
computed easily. With regards to $E(\widehat{\lambda}_{ij})$ we can
use the EM estimates for $\lambda_{ij}$ because they are correct. For
the variances we make use of the procedure described above.

\subsubsection*{The ``Log Ratio'' statistical test}
\ \ \ The variance at the denominator of the statistical test can be
decomposed as $ Var(\ln \widehat{\lambda}_{i1}- \ln
\widehat{\lambda}_{i2})= Var(\ln \widehat{\lambda}_{i1}) + Var(\ln
\widehat{\lambda}_{i2}) $.  Now we observe that $ Var( \ln
\lambda_{ij}) = Var \left(\ln \left(\frac{y_{ij+}}{n_j} \right)\right)
= Var(\ln(y_{ij+})) $. According to the Delta method $
Var(g(y_{ij+})) \approx Var(y_{ij+}) \left(\frac{\partial}{\partial
  y_{ij+}} g(y_{ij+})\right)^2 $ and in our case the Delta
approximation for the variance is $Var(\ln y_{ij+}) =
\frac{1}{y_{ij+}^2} Var(y_{ij+}) $.

\section{Simulation study}\label{sec:sim}

The performance of the proposed strategy is evaluated by a large
simulation study comprising several data generating processes with the
double aim of: (a) assessing the capability of the proposed mixture
model to estimate the variances with a specified number of components,
and (b) evaluating the accuracy of the three statistical test
procedures in terms of power and first-type error. In particular, in a
set of multiple $p$ tests in the absence of correction, a reliable
statistical test should reach the nominal significance level as $n$
increases. We compare the three proposed statistical tests with the
procedures of \cite{robinson2010edger},
\cite{anders2010differential} and \cite{WWW13} and implemented in the packages
\emph{edgeR}, \emph{DESeq} and \emph{DSS} respectively.

\subsection{Simulation A}\label{sec:simA}
In the first simulation study, we evaluated the capability of the
proposed mixture model to estimate the variances of the genes as the
number of components, $K$, increases. Indeed, our purpose is to account for heterogeneity among the overdispersion parameters and we do not necessarily believe that groups of differently overdispersed genes do exist. In the limit, the number of overdispersion parameters could be equal to the number of genes, and we want to test if our strategy will adapt to such a situation.  We also computed some
conventional information criteria in order to select the optimal
number of components.

A set of $H=100$ datasets with $p= 300$ genes, $d=2$ conditions and
$n_j=5$ replicates have been simulated. The $1/3$ of the genes (= 100
genes) are supposed to be differentially expressed ($\lambda_{i1} \neq
\lambda_{i2}$), while the remaining genes (= 200 genes) have been
generated with $\lambda_{i1} = \lambda_{i2}$. For the 100
differentially expressed genes we generated $\lambda_{i1} \sim
Unif(0,250)$ and $ \lambda_{i2} = \frac{\lambda_{i1}}{e^{\phi_i}} $
where $\phi_i$ is randomly drawn from a $N(\mu= 0.5, \sigma= 0.125)$;
for the 200 non-differentially expressed genes we considered
$\lambda_{i1}=\lambda_{i2} \sim Unif(0,250)$. For all the genes the
dispersion parameters have been randomly drawn from a $Unif(0.5,600)$.
All the values for the parameters have been chosen to be consistent
with the empirical situation.

\begin{figure}[t]
\centering
  {\includegraphics[width=0.45\textwidth]{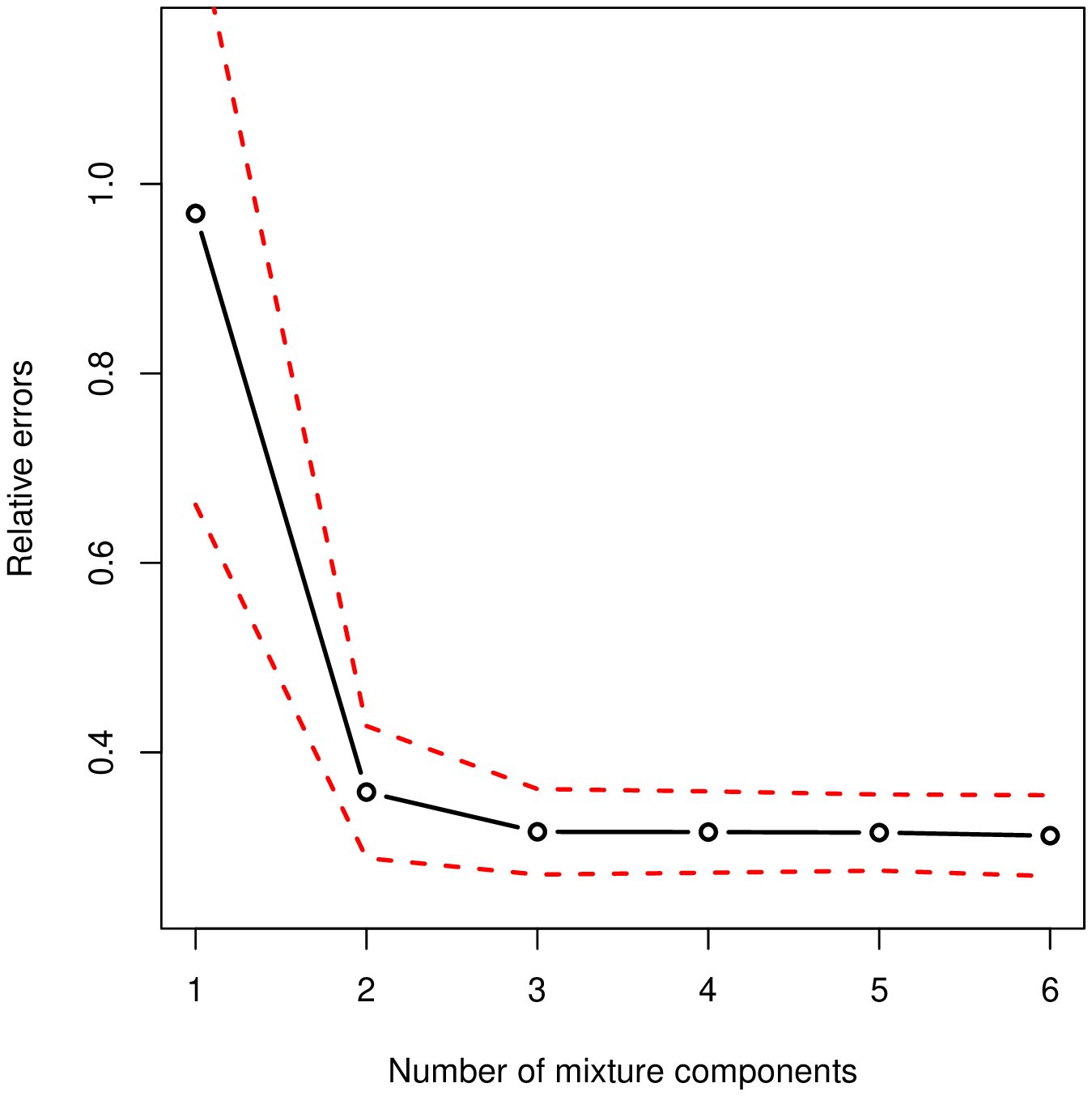}, \
  \includegraphics[width=0.45\textwidth]{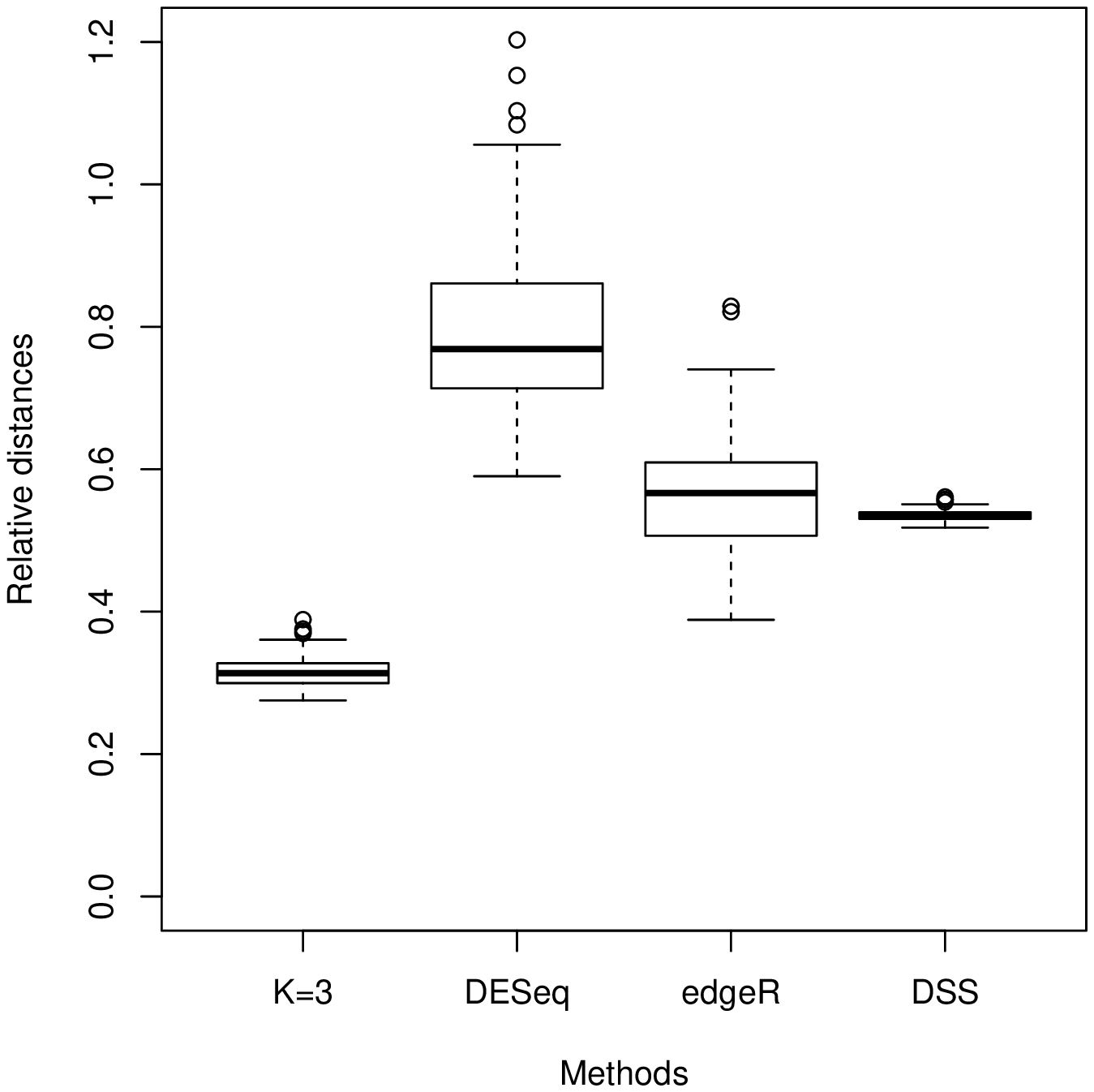}}
\caption{Simulation A: variances estimation. In the first plot, the
  relative distances between the estimated variances and the true ones
  as $K$ varies are shown.  The dashed lines depict the standard error
  bands. In the second graphs, the boxplot of the relative distances
  between the estimated variances and the true ones for all the
  methods.}
\label{simA}
\end{figure}

On each dataset we fitted the proposed mixture model for $K$ ranging
from 1 to 6. Figure \ref{simA} shows the average of the relative
distances in absolute values across the 100 datasets between the
estimated variances by (\ref{var.y}) and the true ones as $K$
varies. In computing the (\ref{var.y}), we applied the correction
factor $n_j/(n_j-1)$ in order to obtain the corresponding correct
estimator. The dashed lines denote the standard error bands computed
as mean $\pm \ 2\cdot$standard error. From the graph, we could say
that from $K=3$ components the gain of fitting more complex mixture
models becomes irrelevant. In other terms, it seems that $K=2$ and
$K=3$ components well describe the variability of the $p$ genes.

This insight is also confirmed by the information criteria. More
specifically, we have considered the Akaike's Information Criterion
(\cite{akaike1974}), $\textrm{AIC}=-2\log \max L+2h$, where $h$ is the
total number of required parameters and the more conservative Bayesian
Information Criterion (\cite{schwarz1978}), $\textrm{BIC}=-2\log\max
L+h\log p$. In addition, we have also computed the so-called
Integrated Classification Likelihood criterion (\cite{biernacki2000})
that combines the BIC penalty term with the entropy of the posterior
classification. As a result, ICL-BIC is characterized by a heavier
penalty term and it tends to favour simpler model against mixture
models with more components.

\begin{table}[b]
\centering
\caption{Simulation A: number of times each information criterion
suggests a specific value for $K$
\label{infoCrit}}
\begin{center}
\begin{tabular}{lrrr}
\hline
 $K$ & AIC & BIC & ICL-BIC\\
\hline
 1 & 0 & 0 & 0 \\
  2 & 0 & 2 & 76 \\
  3 & 76 & 86 & 24 \\
  4 & 6 & 4 & 0 \\
  5 & 6 & 4 & 0 \\
  6 & 12 & 4 & 0 \\
  \hline
\end{tabular}
\end{center}
\end{table}

In Table \ref{infoCrit} the number of times each criterion suggests a
specific number of components $K$ is shown.  These results recommend
that $K=3$ mixture components are enough to give a good description of
the data. The capability of estimating the dispersions and therefore
the variances for each method has been checked through the computation
of the average relative errors between the true variances and the ones
estimated by each analyzed strategies. In Figure \ref{simA} the
relative distances between the estimated and true variances as $k$
varies and the correspondent boxplots are presented. It is clear from
this graph that the proposed method greatly improves the accuracy of
the variance estimation.

\subsection{Simulation B}\label{sec:simB}
\begin{figure}[h]
\centering\includegraphics{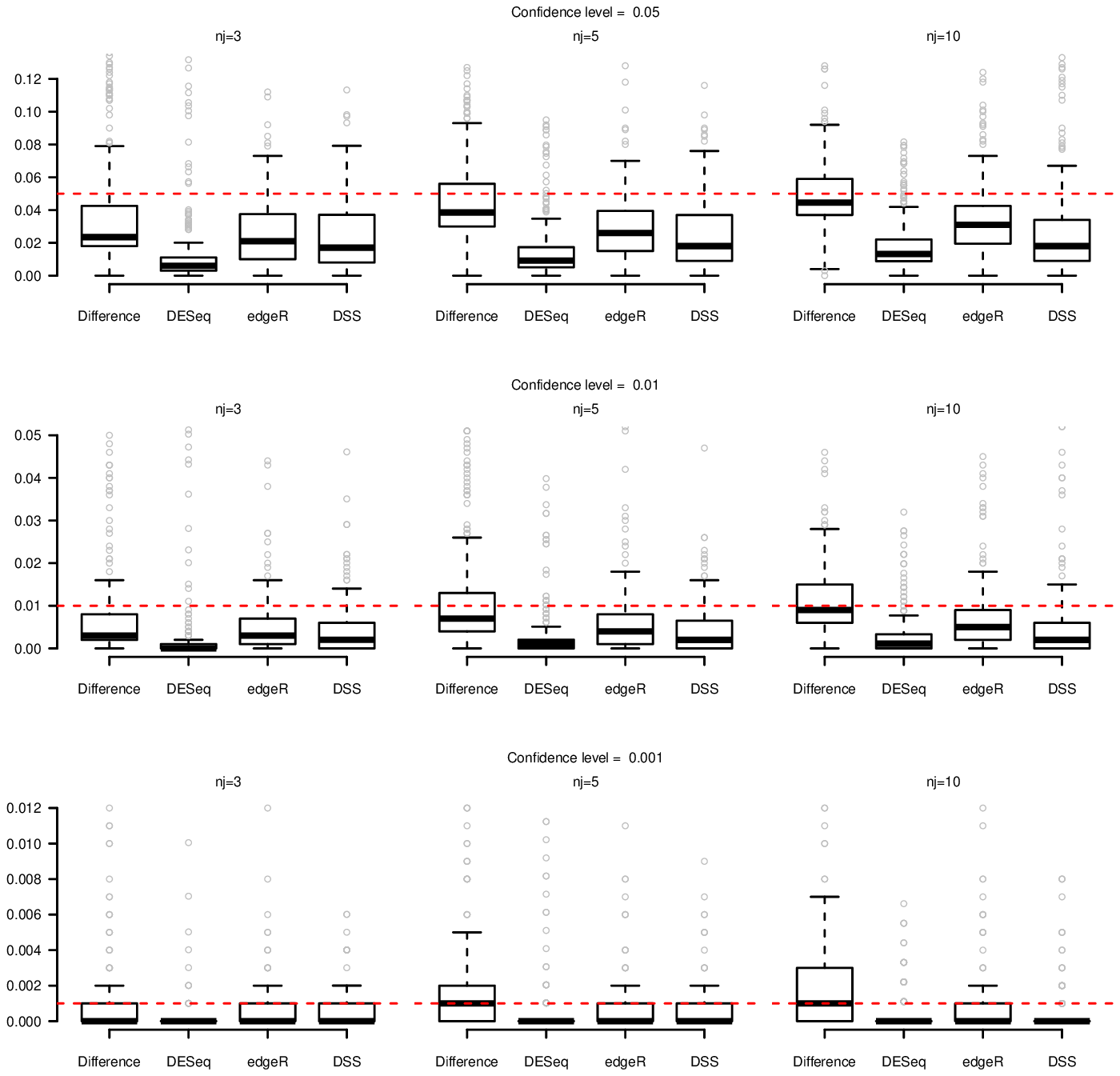}
\caption{Box-plots of the distribution of the first-type errors
  computed on the null genes. Comparison between the performances of
  the proposed ``Difference'' test statistic, DESeq,
  edgeR and DSS as $n_j$
  varies. The dashed line indicates the nominal value that has been
  considered.}
\label{Boxplot1Diff}
\end{figure}

In this simulation study we considered the same simulation design
presented before with a varying number of replicates $n_j= 3, 5, 10$.
For each case we generated $H=1000$ datasets. Then the mixture model
with $K=3$ components has been estimated on the data and the three
proposed test statistics have been computed.  The adequateness of the
statistical procedures can be evaluated by observing the approximation
towards the nominal significance level under the null hypothesis as
the number of replicates increases.  For each of the 200
not-differentially expressed genes, we have computed the empirical
first-type errors across the 1000 datasets for the three statistics.
For comparative purposes, we have also computed the null p-values
provided by \emph{DESeq}, \emph{edgeR} and \emph{DSS} on the same
data. Figure \ref{Boxplot1Diff} contains the box-plots of the
empirical first-type errors obtained by the ``Difference'' test
statistic and of the other considered approaches as the number of
replicates varies and for the different levels of the test (0.05, 0.01
and 0.001).

The three statistical tests fast converge to the nominal level as the
number of the replicates increases, while the \emph{DESeq},
\emph{edgeR} and \emph{DSS} based tests are always under the nominal
level.  It is clear from these graphs that the proposed test
statistics are the only ones that actually reach the nominal value for
the first-type error. The distribution of the first-type errors that
have been obtained from the estimations provided by \emph{edgeR} and
\emph{DSS} crosses the nominal values only with the upper whisker, and
\emph{DESeq} distribution does not cross the nominal value at all.

\begin{figure}[tb]
\centering
\includegraphics{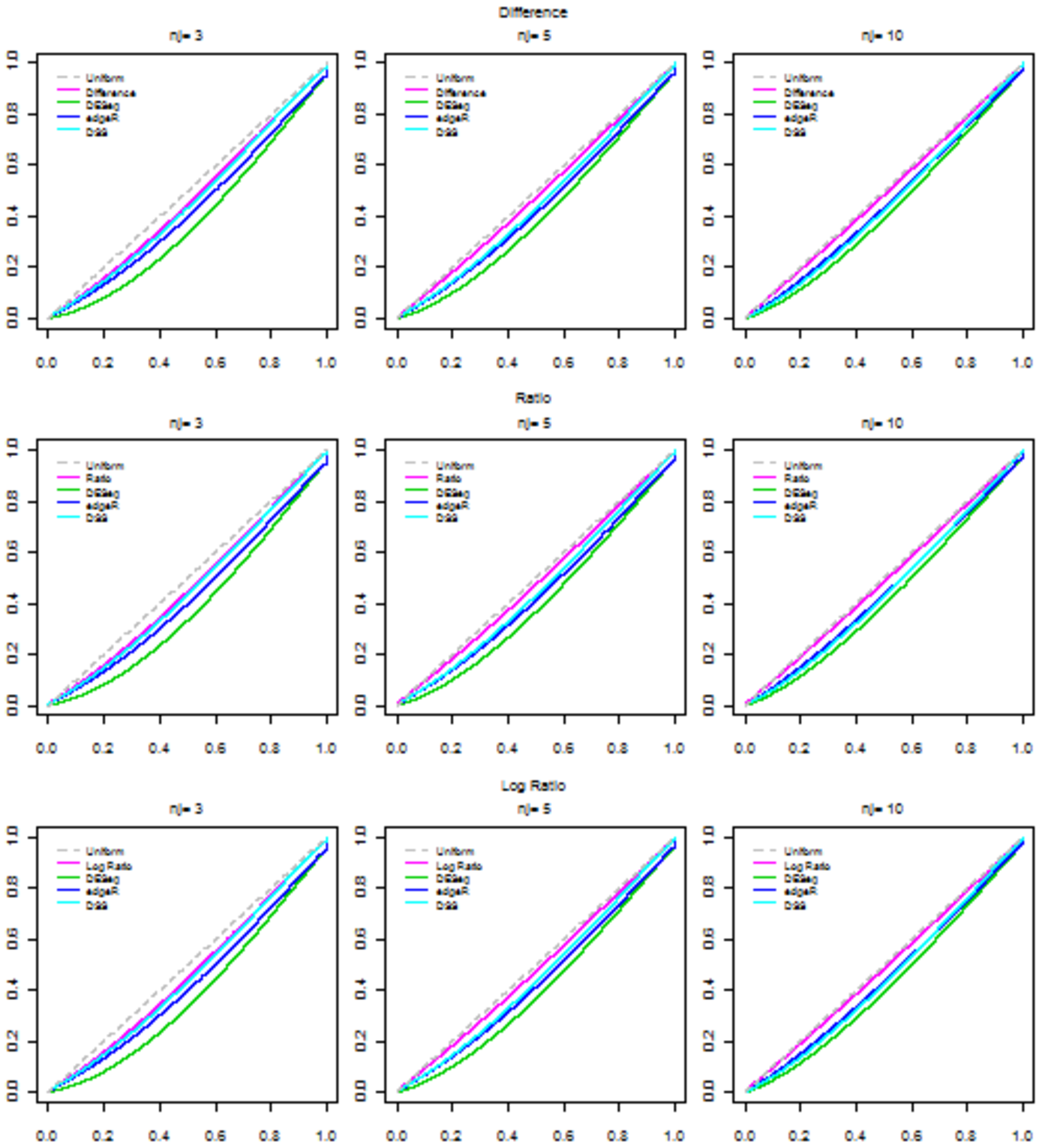}
\caption{Empirical cumulative distribution functions for the null p-values
  that are obtained by the proposed test statistics, DESeq,
  edgeR and DSS. The dashed line indicates the ECDF of the uniform
  distribution, that is the target one.}
\label{ecdf}
\end{figure}

The capability of controlling the first-type error can be checked also
looking at the empirical cumulative distribution function (ECDF) of
the null p-values; the more their distribution is close to the
diagonal, the more they can be considered as actually uniformly
distributed, as requested by the probability integral transform
theorem.  In Figure \ref{ecdf} the ECDFs for the null p-values
obtained through the proposed test statistics, \emph{DESeq},
\emph{edgeR} and \emph{DSS} as $n_j$ varies are shown.  It is clear
that the proposed test statistics behave better than the others
already in correspondence of $n_j= 3$, then the correspondent ECDFs
become closers and closers to the diagonal as the number of replicates
increases and for $n_j= 10$ the ECDF for the null p-values of the
proposed procedures even overlap the diagonal, whereas \emph{edgeR},
\emph{DESeq} and \emph{DSS} reveal curves that lie behind the diagonal
for all the three scenarios.

The means and standard errors of the first-type and second-type errors
have been reported in Tables \ref{err1} and \ref{err2}. It is
important to underline that the distributions of the first- and
second-type errors are very skewed especially with regards to
\emph{DESeq}, \emph{edgeR} and \emph{DSS}. It is possible to
assay that the proposed test procedures (and in particular the
``Difference'' and the ``Log Ratio''), in addition to being able to
control the false-positive rates, are good also in containing the
false-negative ones.

\begin{table}[t]
\centering
\caption{Simulation B: means and standard errors of first-type
  errors at different confidence levels.
\label{err1}}
\begin{tabular}{lrrr}
\hline
 Statistic & $n_j= 3$ & $n_j= 5$ & $n_j= 10$ \\
  \hline \multicolumn{4}{c}{Confidence level=
  0.05}\\ \hline Difference & 0.0392 (0.0356) & 0.0483 (0.0273) &
0.0505 (0.0213) \\ Ratio & 0.0418 (0.0351) & 0.0501 (0.0267) & 0.0516
(0.0211) \\ Log Ratio & 0.0395 (0.0366) & 0.0485 (0.0278) & 0.0506
(0.0217) \\ DESeq & 0.0143 (0.0242) & 0.0172 (0.0206) & 0.0201
(0.0187) \\ edgeR & 0.0337 (0.0454) & 0.0333 (0.0335) & 0.0346
(0.0229) \\ DSS & 0.0380 (0.0624) & 0.0352 (0.0499) & 0.0293 (0.0318)
\\ \hline \multicolumn{4}{c}{Confidence level= 0.01}\\ \hline
Difference & 0.0107 (0.0179) & 0.0121 (0.0134) & 0.0119 (0.0098)
\\ Ratio & 0.0135 (0.0197) & 0.0146 (0.0142) & 0.0131 (0.0104) \\ Log
Ratio & 0.0110 (0.0190) & 0.0123 (0.0138) & 0.0120 (0.0100) \\ DESeq &
0.0036 (0.0111) & 0.0034 (0.0072) & 0.0037 (0.0061) \\ edgeR & 0.0102
(0.0252) & 0.0085 (0.0155) & 0.0074 (0.0085) \\ DSS & 0.0128 (0.0382)
& 0.0102 (0.0260) & 0.0066 (0.0125) \\ \hline
\multicolumn{4}{c}{Confidence level= 0.001}\\ \hline Difference &
0.0031 (0.0086) & 0.0025 (0.0047) & 0.0021 (0.0032) \\ Ratio & 0.0045
(0.0105) & 0.0037 (0.0063) & 0.0026 (0.0039) \\ Log Ratio & 0.0033
(0.0092) & 0.0027 (0.0051) & 0.0021 (0.0034) \\ DESeq & 0.0012
(0.0053) & 0.0007 (0.0023) & 0.0005 (0.0012) \\ edgeR & 0.0032
(0.0126) & 0.0018 (0.0058) & 0.0012 (0.0024) \\ DSS & 0.0048 (0.0211)
& 0.0032 (0.0117) & 0.0013 (0.0038) \\
\hline
\end{tabular}
\end{table}

\begin{table}[h]
\centering
\caption{Simulation B: means and standard errors of second-type errors at different confidence levels.
\label{err2}}
\begin{tabular}{lrrr}
\hline
Statistic & $n_j= 3$ & $n_j= 5$ & $n_j= 10$ \\
\hline
\multicolumn{4}{c}{Confidence level= 0.05}\\
\hline
Difference  & 0.1582 (0.2738) & 0.1002 (0.2267) & 0.0543 (0.1455) \\
Ratio       & 0.2112 (0.3259) & 0.1304 (0.2812) & 0.0764 (0.2046) \\
Log Ratio   & 0.1569 (0.2726) & 0.0991 (0.2246) & 0.0534 (0.1443) \\
DESeq       & 0.1987 (0.3007) & 0.1196 (0.2568) & 0.0642 (0.1809) \\
edgeR       & 0.1444 (0.2526) & 0.0945 (0.2197) & 0.0529 (0.1533) \\
DSS         & 0.1354 (0.2449) & 0.0892 (0.2109) & 0.0513 (0.1526) \\
\hline
\multicolumn{4}{c}{Confidence level= 0.01}\\
\hline
Difference  & 0.2341 (0.3289) & 0.1442 (0.2867) & 0.0874 (0.2199) \\
Ratio       & 0.3336 (0.3874) & 0.1897 (0.3334) & 0.1146 (0.2775) \\
Log Ratio   & 0.2331 (0.3278) & 0.1430 (0.2845) & 0.0856 (0.2167) \\
DESeq       & 0.3141 (0.3472) & 0.1755 (0.3102) & 0.0980 (0.2462) \\
edgeR       & 0.2268 (0.2997) & 0.1384 (0.2740) & 0.0815 (0.2170) \\
DSS         & 0.2159 (0.3014) & 0.1357 (0.2710) & 0.0813 (0.2181) \\
\hline
\multicolumn{4}{c}{Confidence level= 0.001}\\
\hline
Difference  & 0.3441 (0.3703) & 0.2037 (0.3345) & 0.1228 (0.2834) \\
Ratio       & 0.5075 (0.3996) & 0.2889 (0.3847) & 0.1545 (0.3260) \\
Log Ratio   & 0.3433 (0.3693) & 0.2026 (0.3333) & 0.1212 (0.2799) \\
DESeq       & 0.4873 (0.3635) & 0.2620 (0.3572) & 0.1382 (0.3016) \\
edgeR       & 0.3609 (0.3359) & 0.2066 (0.3193) & 0.1166 (0.2753) \\
DSS         & 0.3508 (0.3471) & 0.2061 (0.3230) & 0.1176 (0.2758) \\
\hline
\end{tabular}
\end{table}

\section{Application to Prostate Cancer Data}\label{sec:appl}

We have analyzed data on RNA-Seq data on prostate cancer cells collected in two different conditions: a group of patients has been treated with androgens, and the
second one with an inactive compound. The data have been sequenced and analyzed by \cite{Li2008}. It is well known that androgen hormones stimulate some genes,
and they also have a positive effect in curing prostate cancer cells. Therefore the connection between these stimulated genes and the survival of these cells is a
largely studied issue.
Seven biological replicates of prostate cancer cells (three for the androgen-treated condition and four for the control-group) for $ 37435 $ genes have been
sequenced using the Illumina 1G Genome Analyzer. Then they have been mapped to the NCBI36 build of the human genome using Bowtie (allowing up to two mismatches)
and then the number of reads that corresponded to each Ensembl gene (version 53) was counted. The resulting read count table is available from
\verb"https://sites.google.com/site/davismcc/useful-documents". For the analysis we have considered the $p= 16424$ genes with mean count greater than 1, because
they provide sufficient statistical information on the differential analysis. In order to account for the bias introduced by the different lanes of the experiment and the eventual effect the gene length, we preliminarily normalized the data using quantile-based normalization scheme implemented in the R package \emph{EDASeq} (see, among the others, \cite{risso2011}, \cite{TGD11}, \cite{DRA13}, \cite{BPH10}).

\subsection{Analysis and Results}
The proposed NB mixture model has been fitted on the data with a
number of components $K$ ranging from 1 to 6. The BIC and AIC criteria
suggested $K=3$ components. Convergence has been obtained with 50
iterations of the EM-algorithm at the log-likelihood of -384886 (BIC=
1088660, AIC= 835479).  Differential expression analysis has been
conducted by computing the three proposed test statistics. For
comparative purposes we have performed differential analysis using the
\emph{DESeq}, \emph{edgeR} and \emph{DSS} methods implemented in
\verb"R" using the default settings. The estimation of the dispersion
parameters have been obtained with the following R commands:
\verb"estimateDispersions" for \emph{DESeq},
\verb"estimateTagwiseDisp" for \emph{edgeR} and \verb"estDispersion"
for \emph{DSS}.

All the obtained p-values have been adjusted following the procedure
of \cite{benjamini1995} in order keep under control the total first
error in multiple comparison testing.  In Table \ref{RealA} the number
of genes declared DE by each method at the confidence levels of 0.05,
0.01 and 0.001 is shown. The different methods detect a proportion of
DE genes ranging from about 10$\%$ to 25$\%$. In order to investigate
the degree of accordance between two methods, we measured the
proportion between the number of genes declared DE jointly by both
methods and the average number of the genes declared DE marginally at
a certain confidence level.

\begin{table}[h]
\centering
\caption{Number of genes declared DE for all the compared methods at different confidence levels (adjusted p- values)
\label{RealA}}
\begin{tabular}{lrrr}
\hline {Statistic} & $\alpha= 0.05$ & $\alpha= 0.01$ & $\alpha= 0.001$ \\
\hline
Difference  & 3167 & 2146 & 1360 \\
Ratio       & 3538 & 2591 & 1914 \\
Log Ratio   & 4254 & 2941 & 2024 \\
DESeq       & 2695 & 1828 & 1271 \\
edgeR       & 3918 & 2774 & 1886 \\
DSS         & 4215 & 2737 & 1737 \\
\hline
\end{tabular}
\end{table}

\begin{figure}[bh]
\centering
{  \includegraphics[width=0.32\textwidth]{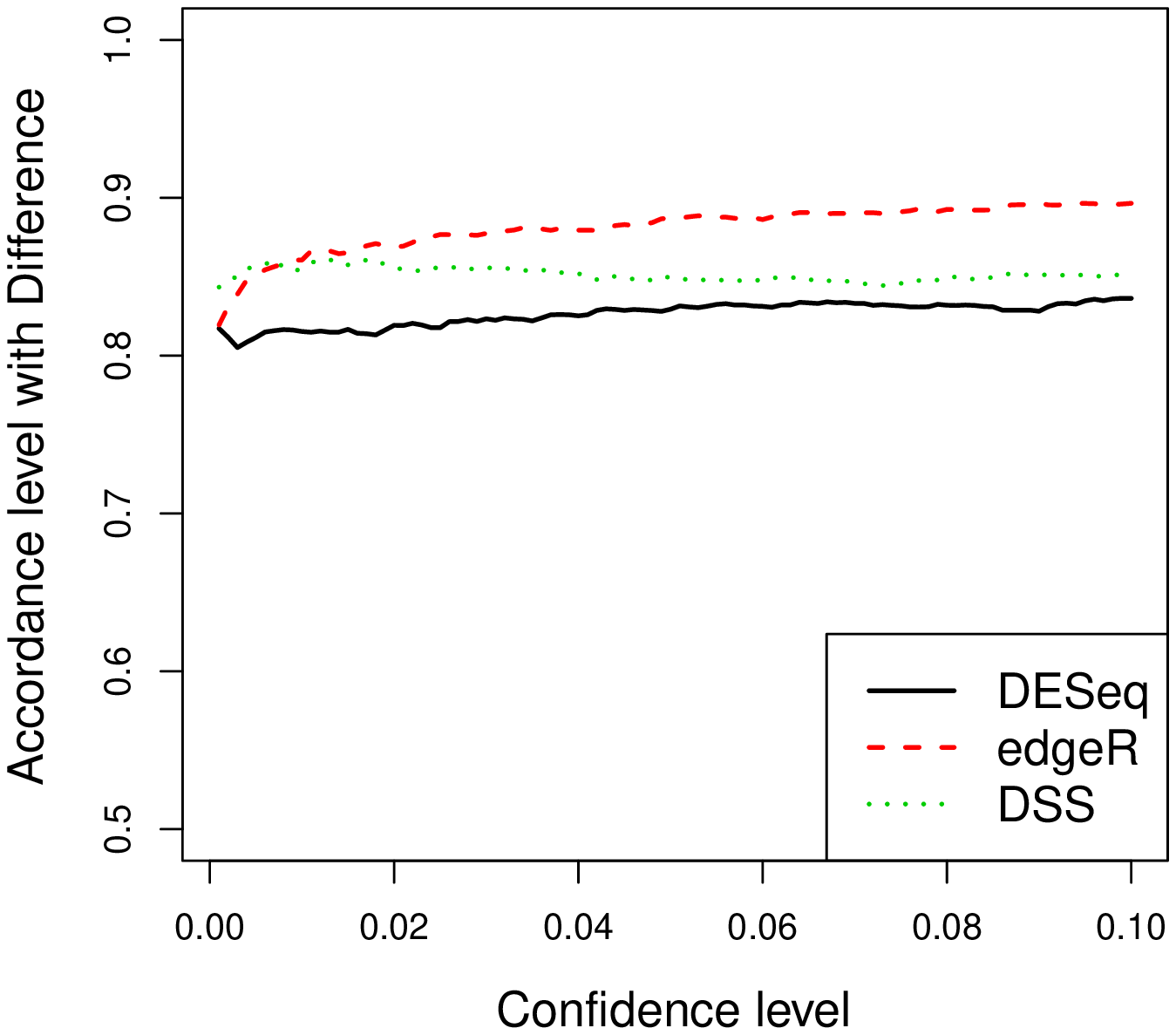},  \includegraphics[width=0.32\textwidth]{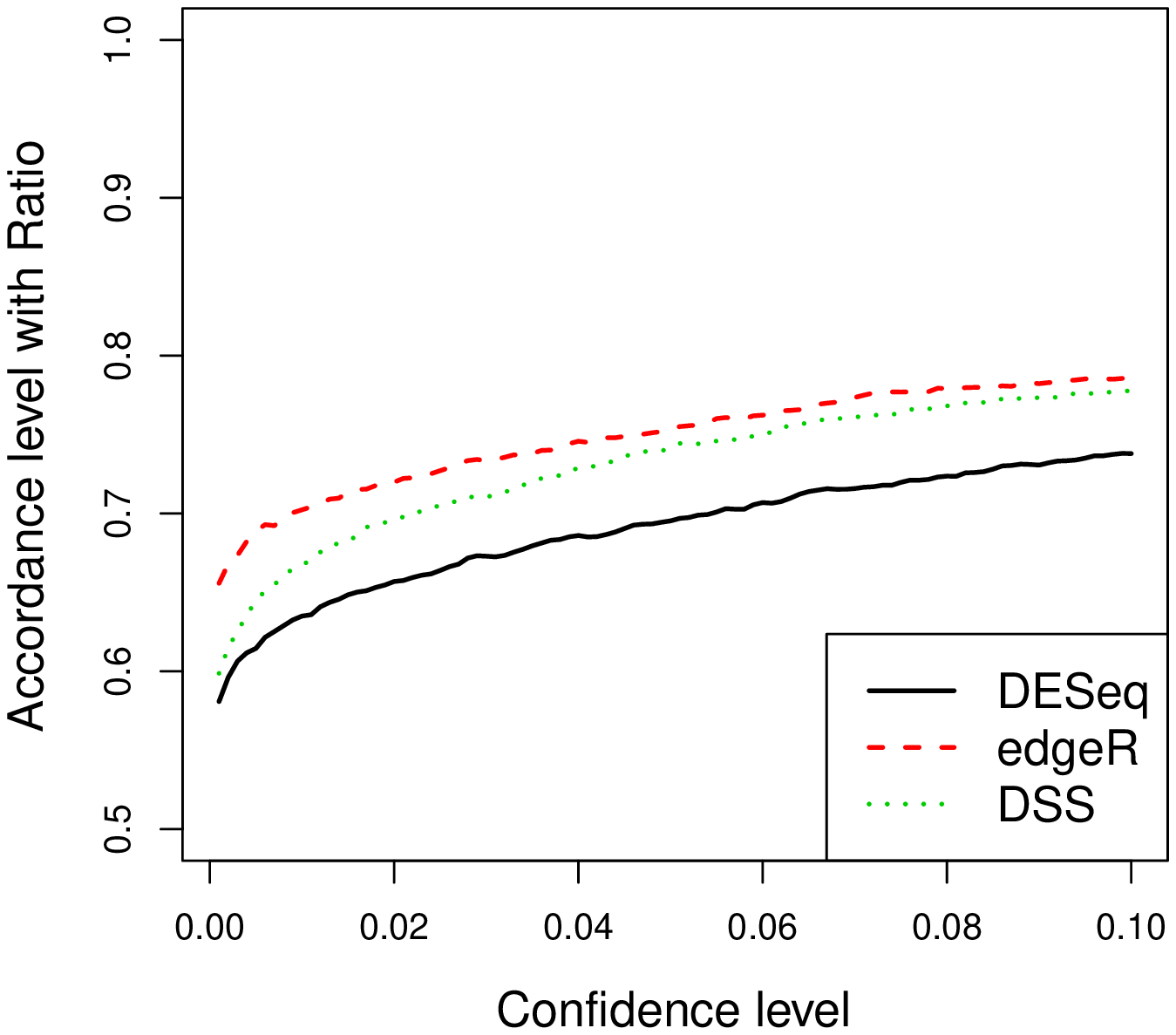},
  \includegraphics[width=0.32\textwidth]{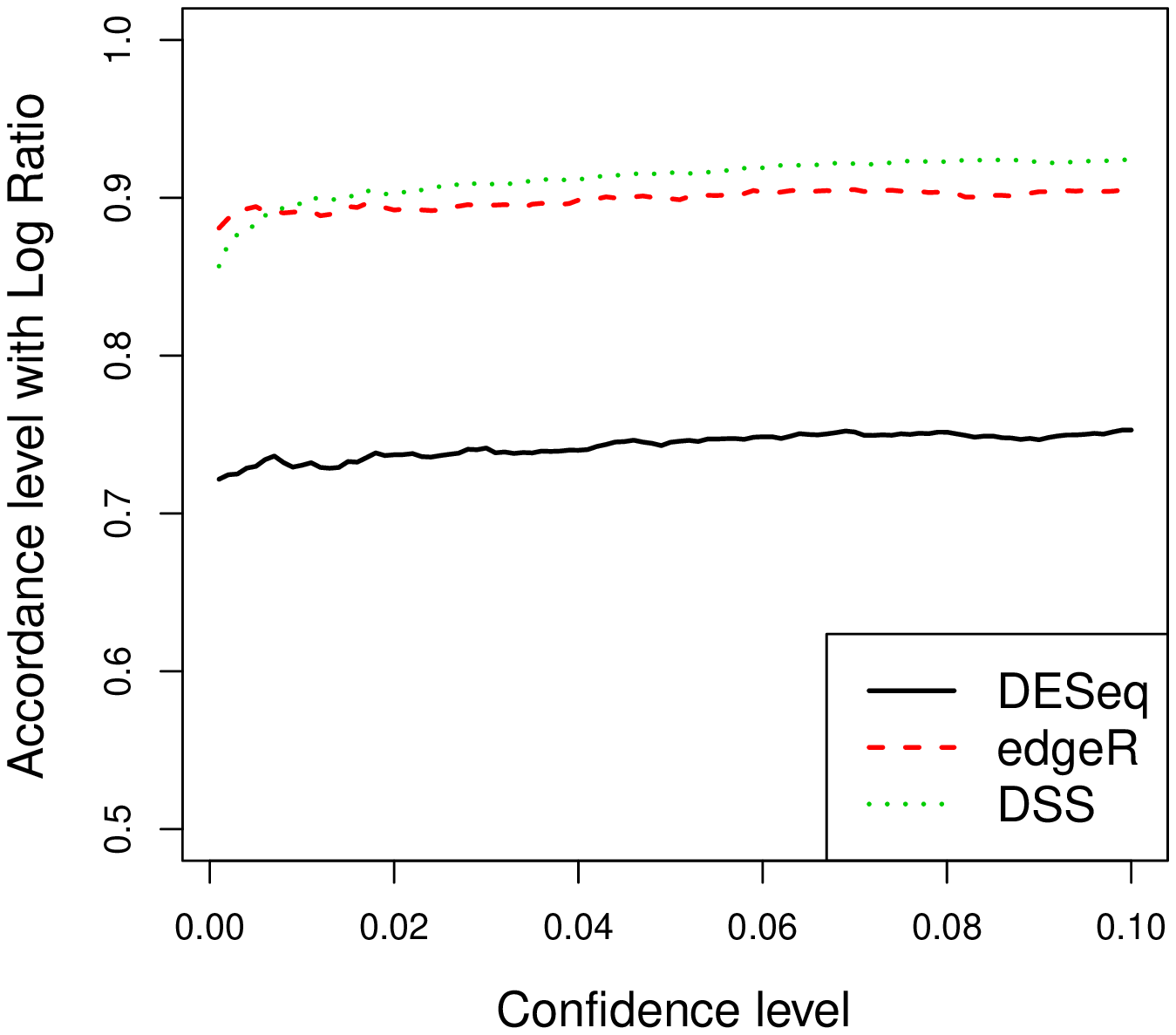}}
\caption{Proportion of genes declared DE as the confidence level
  increases for the different methods where the proposed
  ``Difference'' (panel a), ``Ratio'' (panel b) and ``Log Ratio'' test
  statistic (panel c) are taken as baseline.}
\label{Proplev}
\end{figure}

The first panel of Figure \ref{Proplev} shows the pairwise comparison
between the proposed ``Difference'' test statistic and the
\emph{DESeq}, \emph{edgeR} and \emph{DSS} methods. The other two
pictures of Figure \ref{Proplev} show the same results for the
``Ratio'' and ``Log Ratio'' test statistics respectively.

It is clear from these graphs that the proposed test statistics
provide results that are strongly consistent with the ones obtained by
\emph{edgeR} and \emph{DSS}, with a degree of accordance of about
90$\%$ when the ``Difference'' or the ``Log Ratio'' is used.  The set
of DE genes detected by \emph{DESeq} seems to be slightly different by
the ones selected by all the other methods, even if the accordance
level is between 60$\%$ and 80$\%$.

\section{Concluding remarks}\label{sec:dis}

We proposed a novel framework for the differential analysis of count data in the negative binomial setting, especially designed for the analysis of RNA-Seq data. Like several others already proposed, our approach accounts for the heterogeneity of the overdispersion parameter across genes, but the use of a mixture model to the aim is novel. Our approach is fully consistent in terms of parameter estimation and hypothesis testing. As a result, the first-type error of the proposed test is controlled.

The comparative study we performed shows that the proposed strategy is competitive with existing methods. It also shows that some popular testing procedures like \emph{DEseq} or \emph{edgeR} actually do not control the first-type error. This lack of control is likely to be due to the post-processing of the overdispersion parameter, which is not accounted for by the null-distribution of the test statistics. The control achieved by \emph{DSS}, which also combines consistent parameter estimation and testing methods, is similar to ours.

In this paper, we focus on two sample comparison, but the procedure can indeed be adapted to any contrast, in an obvious manner, especially when using the ``Difference'' statistics. In a similar way, because our approach can be cast in the general linear framework, normalization or correction for some exogenous effects could also be considered.

\bibliographystyle{plainnat}
\bibliography{biblio_art}

\appendix


\section{Appendix}
\subsection{$\textbf{Y}_{i}$ follows a mixture of NB
distributions}
If $Y_{ijr}|U_{ijr}=u_{ijr} \sim Poisson(\lambda_{ij}u_{ijr})$ then $\textbf{Y}_{i}$ follows a mixture of NB
distributions.

PROOF \\

Without loss of generality, we drop from the proof the subscripts denoting the replicates. The proof can be obtained as follows:
\begin{equation*}
\begin{split}
    f(y) = & \int_0^{+\infty} f(y,u) du \\
     = & \int_0^{+\infty} f(y|u) f(u) du \\
     = & \int_0^{+\infty} \frac{e^{-(\lambda u)} (\lambda u)^{y}}{y!} \sum_k w_k \frac{\alpha_k^{\alpha_k}}{\Gamma (\alpha_k)} u^{\alpha_k-1} e^{-\alpha_k u} du \\
     = & \frac{1}{y!} \lambda^y \sum_k w_k \frac{\alpha_k^{\alpha_k}}{\Gamma (\alpha_k)} \int_0^{+\infty} u^y u^{\alpha_k-1}e^{-\alpha_k u}e^{- \lambda u} du\\
     = & \sum_k w_k \frac{\lambda^y \alpha_k^{\alpha_k}} {\Gamma(y+1) \Gamma(\alpha_k)} \underbrace{\int_0^{+ \infty} u^{y+\alpha_k-1} e^{-u(\alpha_k + \lambda)} du}_{\mbox{kernel of a $Gamma(y + \alpha_k, \alpha_k + \lambda) $}}\\
     = & \sum_k w_k \frac{\lambda^y \alpha_k^{\alpha_k}} {\Gamma(y+1) \Gamma(\alpha_k)} \frac{\Gamma(y+ \alpha_k)}{(\lambda + \alpha_k)^{y+\alpha_k}}\\
     = & \sum_k w_k \binom{y+\alpha_k-1}{\alpha_k-1}  \left( \frac{\lambda}{\lambda + \alpha_k} \right)^y \left( \frac{\alpha_k}{\lambda + \alpha_k} \right )^{\alpha_k}
\end{split}
\end{equation*}

    \subsection{EM algorithm}
In order to develop the Expectation and Maximization steps, we expand the conditional expectation in \ref{EM} as the
sum of three terms:

\begin{eqnarray}\label{El_c}
&& \ \ \ E_{\textbf{z},\textbf{u}|\textbf{y};
    \boldsymbol\Theta'}\left[ \log
    f(\textbf{y},\textbf{u},\textbf{z}|\boldsymbol\Theta)\right]=\int
  \sum_{k=1}^K \sum_{i=1}^p \sum_{j=1}^d \sum_{r=1}^{n_j} \ln
  f(y_{ijr}| u_{ijr}; \boldsymbol\Theta) f(
  u_{ijr},\textbf{z}_i|\textbf{y}_i; \boldsymbol\Theta') du_{ijr}+
  \nonumber \\ &+& \int \sum_k \sum_i \sum_j \sum_r \ln
  f(u_{ijr}|\textbf{z}_i; \boldsymbol\Theta) f(
  u_{ijr},\textbf{z}_i|\textbf{y}_i \boldsymbol\Theta') du_{ijr}+
  \sum_k \sum_i\ln f(\textbf{z}_i|\boldsymbol\Theta)
  f(\textbf{z}_i|\textbf{y}_i; \boldsymbol\Theta').
\end{eqnarray}

\subsubsection*{E-step} \ \ \
In the E step we need to compute the conditional densities
$f(u_{ijr}|\textbf{z}_i,\textbf{y}_i)$, $f(\textbf{z}_i|\textbf{y}_i)$
and $f(u_{ijr}|\textbf{y}_i)$ given the current parameter estimates.

The conditional density $f(u_{ijr}|\textbf{z}_i,\textbf{y}_i)$ can be
computed as follows:

\begin{eqnarray}\label{fu.yz}
	 && f(u_{ijr}|\textbf{z}_i,\textbf{y}_i) =
  \frac{f(\textbf{y}_i|u_{ijr},\textbf{z}_i)
    f(u_{ijr}|\textbf{z}_i)f(\textbf{z}_i)}{f(\textbf{y}_i|\textbf{z}_i)
    f(\textbf{z}_i)} = \frac{f(u_{ijr}|\textbf{z}_i) \prod_j \prod_r
    f(y_{ijr}|u_{ijr})}{\prod_j \prod_r f(y_{ijr}|\textbf{z}_i)}
  \nonumber \\ & \propto& \prod_j \prod_r
  e^{-u_{ijr}\lambda_{ij}}(\lambda_{ij}u_{ijr})^{y_{ijr}}u_{ijr}^{\alpha_k-1}e^{-\alpha_ku_{ijr}}
  = \prod_j \prod_r e^{-u_{ijr}(\lambda_{ij}+ \alpha_k)}
  u_{ijr}^{\alpha_k - 1 + y_{ijr}}
 \end{eqnarray}
and since we are considering just the single $u_{ijr}$, we
recognize the probability density function (pdf) of a $ Gamma(y_{ijr} + \alpha_k
, \lambda_{ij}+ \alpha_k) $ since all the factors of the products that
concerned $i'\neq i$ and $j'\neq j $ can be viewed as constant
terms. This is a direct consequence of the fact that the Gamma distribution is the conjugate of the Poisson distribution (applied
  conditionally to the group $z_i$).

The conditional density $f(\textbf{z}_i|\textbf{y}_i)$ can be derived
by the Bayes' rule:
\begin{eqnarray}\label{fz.y}
f(\textbf{z}_i|\textbf{y}_i) = \frac{w_k \prod_j \prod_r
  f(y_{ijr}|z_{ik}=1)}{\sum_k w_k \prod_j \prod_r f(y_{ijr}|z_{ik}=1)}
,
 \end{eqnarray}
 and finally the density $f(u_{ijr}|\textbf{y}_i)$ can be obtained by
 the previous two posteriors
  as follows:
 \begin{eqnarray}\label{fu.y}
 f(u_{ijr}|\textbf{y}_i)=\sum_{k=1}^K
 f(u_{ijr},z_{ik}|\textbf{y}_i)=\sum_{k=1}^K
 f(u_{ijr}|z_{ik},\textbf{y}_i)f(z_{ik}|\textbf{y}_i)
  \end{eqnarray}

\subsubsection*{M-step} 	\ \ \
Given the previous posterior distributions, the maximum likelihood for
the model parameters can be obtained by evaluating the score function
of (\ref{El_c}) at zero, with respect to each parameter of the model. \\
For the estimation of $\lambda_{ij}$ we can focus on the first term of
(\ref{El_c}) given that it is the only one addend that involves the
parameters $\lambda_{ij}$:

\begin{equation}
\begin{split}
            & \frac{\partial}{\partial \lambda_{ij}}
  \int_0^{+\infty}\sum_{k=1}^K \sum_{i=1}^p \sum_{j=1}^d
  \sum_{r=1}^{n_j} \ln f(y_{ijr}| u_{ijr})f(
  u_{ijr},\textbf{z}_i|\textbf{y}_i) du_{ijr} \\ & =
  \frac{\partial}{\partial \lambda_{ij}} \int_0^{+\infty}\sum_k \sum_i
  \sum_j \sum_r \ln \underbrace{f(y_{ijr}|
    u_{ijr})}_{Pois(\lambda_{ij}u_{ijr})} f( u_{ijr}|\textbf{y}_i,
  \textbf{z}_i) f(\textbf{z}_i|\textbf{y}_i) du_{ijr} \\ & =
  \int_0^{+\infty}\sum_k \sum_r \left( -u_{ijr} +
  \frac{y_{ijr}}{\lambda_{ij}} \right)
  f(u_{ijr}|y_{ijr},\textbf{z}_i)f(\textbf{z}_i|\textbf{y}_i) du_{ijr}
  \\ & = \sum_r \frac{y_{ijr}}{\lambda_{ij}} - \sum_k \sum_r
  E(u_{ijr}|y_{ijr},\textbf{z}_i)f(\textbf{z}_i|\textbf{y}_i)
\end{split}
\end{equation}
and by evaluating it at 0 we get $ \widehat{\lambda_{ij}}=
\frac{\sum_r y_{ijr}}{\sum_k f(\textbf{z}_i|\textbf{y}_i) \sum_r
  E(u_{ijr}|\textbf{y}_i,\textbf{z}_i)} \mbox{,} $ where it can be
proved that the
denominator $\sum_k f(\textbf{z}_i|\textbf{y}_i) \sum_r
E(u_{ijr}|\textbf{y}_i,\textbf{z}_i)$ is simply equal to $n_j$:

PROOF\\

   We set:
  \begin{equation*}
   \delta_{ij} = \sum_{k=1}^K f(\textbf{z}_i|\textbf{y}_i)
   \sum_{r=1}^{n_j} E(u_{ijr}|\textbf{y}_i,\textbf{z}_i)
  \end{equation*}
  and for the (\ref{fu.yz})
  \begin{equation*}
   \delta_{ij} = \sum_{k=1}^K f(\textbf{z}_i|\textbf{y}_i)
   \sum_{r=1}^{n_j} \frac{y_{ijr}+\alpha_k}{\lambda_{ij}+\alpha_k}
  \end{equation*}

  We note that we are considering one specific gene $i$ in the
  condition $j$, and given that the mixture structure involves the
  gene level, $y_{ijr}$ with $r= 1, ..., n_j$ can be considered as
  independently distributed according to a negative binomial, with
  dispersion parameter depending on the group membership of the gene
  $i$ to the $k$-th component of the mixture; therefore
  $f(\textbf{z}_i|\textbf{y}_i)$ is equal to 1 in correspondence to
  the $k-th$ group at which the gene $i$ belongs, and 0 otherwise:
    \begin{equation*}
   \delta_{ij} = \sum_{r=1}^{n_j} \frac{y_{ijr}+\alpha_k}{\lambda_{ij}+\alpha_k}
   \end{equation*}
  therefore:
   \begin{equation*}
  \sum_{r=1}^{n_j} (y_{ijr}+\alpha_k)=  \delta_{ij} (\lambda_{ij}+\alpha_k)
  \end{equation*}
   \begin{equation*}
  \sum_{r=1}^{n_j} y_{ijr} + n_j \alpha_k = \delta_{ij} \lambda_{ij} + \delta_j \alpha_k
  \end{equation*}
  but since $\sum_{r=1}^{n_j} y_{ijr}=\delta_j \lambda_{ij} $,  $n_j \alpha_k$ must be equal to $\delta_{ij} \alpha_k$
  and $n_j = \delta_{ij}$ .

 Therefore

\begin{equation}\label{stimalambda}
\widehat{\lambda_{ij}}= \frac{\sum_r y_{ijr}}{n_j} \mbox{.}
\end{equation}

With regards to $\alpha_k$, we evaluate the score of the second term
of (\ref{El_c}). The estimates for $\alpha_k$ are not in closed-form
since
\begin{eqnarray*}
         && \frac{\partial}{\partial \alpha_k}
        \int_0^{+\infty}\sum_{k=1}^K \sum_{i=1}^p \sum_{j=1}^d
        \sum_{r=1}^{n_j} \ln f(u_{ijr}|\textbf{z}_i)f(
        u_{ijr},\textbf{z}_i|\textbf{y}_i; \theta^{(h-1)}) du_{ijr}
        \\ &=& \frac{\partial}{\partial \alpha_k}
        \int_0^{+\infty}\sum_{k=1}^K \sum_{i=1}^p \sum_{j=1}^d
        \sum_{r=1}^{n_j}(\alpha_k\ln
        \alpha_k + (\alpha_k-1)\ln u_{ijr} - \alpha_ku_{ijr} \\ && - \ln
        \Gamma(\alpha_k))
        f(u_{ijr}|y_{ijr},\textbf{z}_i)f(\textbf{z}_i|\textbf{y}_i) d
        u_{ijr} \\
        & =& \alpha_k \ln \alpha_k - \ln \Gamma(\alpha_k) +
        (\alpha_k-1) E(\ln u_{ijr} | y_{ijr}, \textbf{z}_i) - \alpha_k
        E(u_{ijr}|y_{ijr},\textbf{z}_i) f(\textbf{z}_i|\textbf{y}_i),
\end{eqnarray*}
but they can be obtained by the quasi-Newton algorithm.

With regards to $E(\ln u_{ijr} | y_{ijr}, \textbf{z}_i)$ we can use an already-known result
that states that, given a random variable $X\sim Gamma(\alpha,\beta)$,
$E(\ln X)= \psi(\alpha) - \ln(\beta) $ where  $\psi$ is the digamma function.
Thus, for the (\ref{fu.yz}) we have:
$ E(\ln u_{ijr} | y_{ijr}, \textbf{z}_i)= \psi(y_{ijr} + \alpha_k) - \ln(\lambda_{ij} + \alpha_k) \mbox{.} $

Finally, the maximum likelihood estimate for $w_k$ we be obtained by
maximizing the third term of (\ref{El_c}), from which we obtain:
$$ \widehat{w_{k}}= \frac{\sum_i f(\textbf{z}_i|\textbf{y}_i)}{p} \mbox{.} $$

\end{document}